# The economic, social & Environmental impact of Electric Vehicle (EV) adaptation on Bangladeshi Society.

**Abdullah Al Noman**

**Computer Science & Engineering, North South University, Dhaka-1229, Bangladesh**

Email: abdullah.noman02@northsouth.edu

**Hasibul Hassan Siam**

**Electrical and Electronics Engineering, North South University, Dhaka- 1229, Bangladesh**

Email: hasibul.siam@northsouth.edu

# Table of Contents

# Introduction

Growing concerns about climate change, air pollution, and the decreasing supply of fossil fuel resources have increased the global shift towards sustainable transportation in recent years. Electric vehicles (EVs), which are powered entirely or partially by electricity, have become a competitive alternative to traditional cars with internal combustion engines. To lower greenhouse gas (GHG) emissions, increase energy efficiency, and create cleaner urban environments, several developed and developing nations are implementing EV technologies.

Bangladesh, a South Asian country that is developing quickly and seeing a rise in urbanization, has serious transportation-related issues, such as high air pollution, heavy traffic, and growing fuel import prices. In this regard, EV adoption offers the nation's social progress, economic expansion, and environmental sustainability both opportunities and problems. Although Bangladesh's EV penetration rate is still low by international standards, government programs, private sector investments, and technology developments show that interest in this market is growing.

Studying the impacts of Electric Vehicle (EV) adoption in Bangladesh is important for shaping effective policies and future planning. Economically, EVs can lower running costs — 93% of surveyed respondents believe EVs are cheaper to operate than fuel-based vehicles — and create jobs in areas such as charging services and battery maintenance, though they require significant infrastructure investment [1]. Socially, EVs can improve mobility for low-income groups (75.4% agreement) and reduce urban air pollution, leading to better public health [2]. Environmentally, EVs can cut harmful emissions — 82.5% of respondents agreed they help reduce pollution — supporting Bangladesh's climate goals [3]. Understanding these impacts will help ensure EV adoption is affordable, inclusive, and sustainable

The primary objective of this research is to comprehensively evaluate the economic, social, and environmental impacts of EV adoption in Bangladesh, identifying both potential benefits and barriers.

# Research Methodology

This study employs a mixed-methods quantitative approach, combining a primary quantitative survey with secondary numerical data from published reports and academic studies.In order to connect the primary findings with existing research, the methodology is intended to capture current awareness, usage patterns, attitudes, and socioeconomic aspects relevant to the adoption of electric vehicles (EVs) in Bangladesh.

A quantitative survey of 57 respondents (convenience sampling) was conducted via digital questionnaires. Questions covered:

**Awareness/knowledge:** EV types, technology familiarity.

**Usage patterns:** Frequency, vehicle preferences.

**Attitudes:** Benefits, barriers, policy views.

**Socioeconomic factors:** Income accessibility, job creation.

Secondary data from government reports and academic studies supplemented survey findings.

# Limitations of this Research

The limitations of this study may impact the generalisability of the findings. First off, there were only 57 participants in the sample, which is insufficient to adequately represent all groups in Bangladesh. Second, the majority of responses were from cities, where EVs are more likely to be available and known about. This implies that the opinions of those living in rural regions, where circumstances differ, might not be accurately represented. Third, due to time constraints, the survey was unable to reach a larger and more diverse sample size or run for an extended period of time. Lastly, certain crucial information from industry or government sources was either unavailable or insufficiently detailed to support a more thorough investigation. It is important to remember these things when reading this paper.

## Research Ethics

To protect participants, this study complied with fundamental ethical guidelines. All participants provided their informed consent before the survey began, indicating that they understood the goal of the study before agreeing to participate. Responses were anonymous since their names and other personal information were not gathered. No individual could be recognised from their responses because all findings were displayed in aggregated data, or group form. The information has been used only for this study, and all results were presented truthfully and unaltered.

# Findings

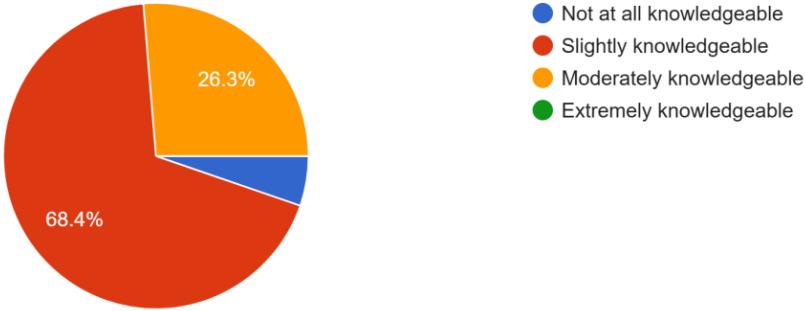

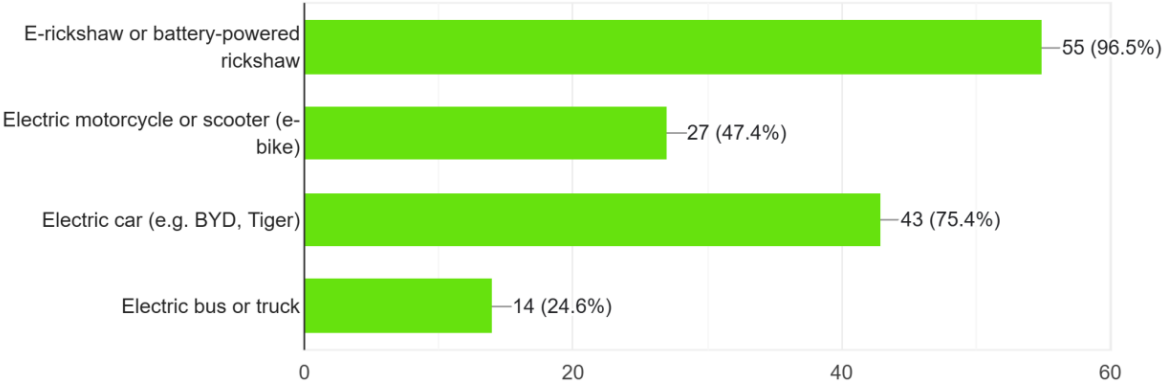

How often do you personally use electric vehicles for transportation (for example, riding in an e-rickshaw or electric car )?
57 responses

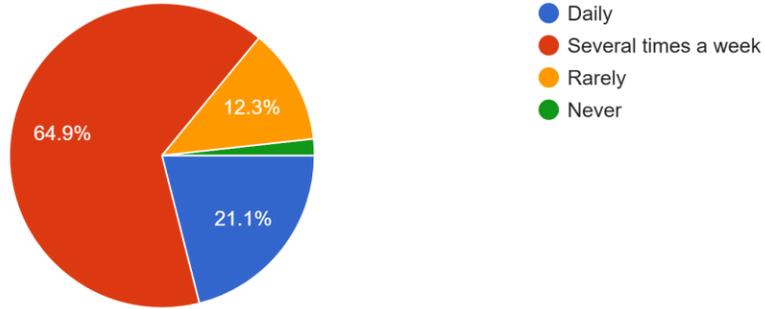

- Daily: 21.1%
- Several times a week: 64.9%
- Rarely: 12.3%
- Never

If you have used an EV, which type(s) have you used? (Select all that apply.)
57 responses

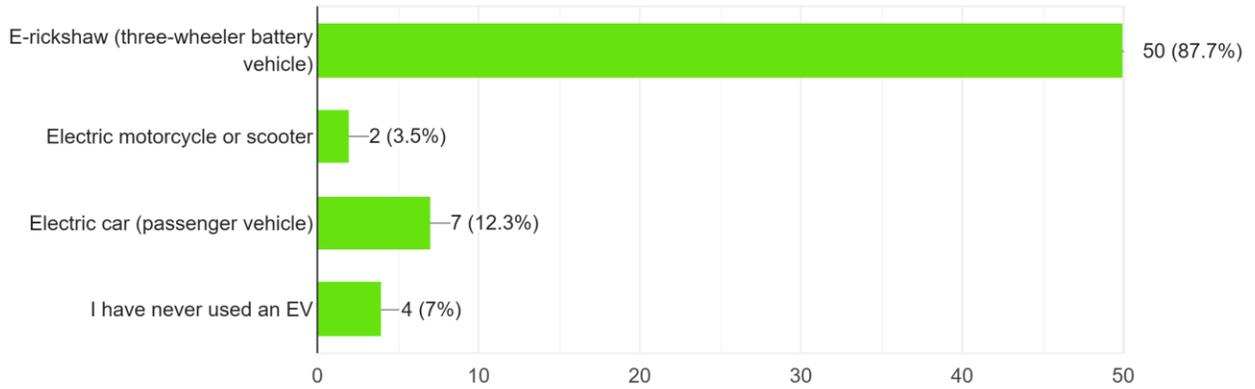

- E-rickshaw (three-wheeler battery vehicle): 50 (87.7%)
- Electric motorcycle or scooter: 2 (3.5%)
- Electric car (passenger vehicle): 7 (12.3%)
- I have never used an EV: 4 (7%)

Are you considering buying or using an electric vehicle in the future?
57 responses

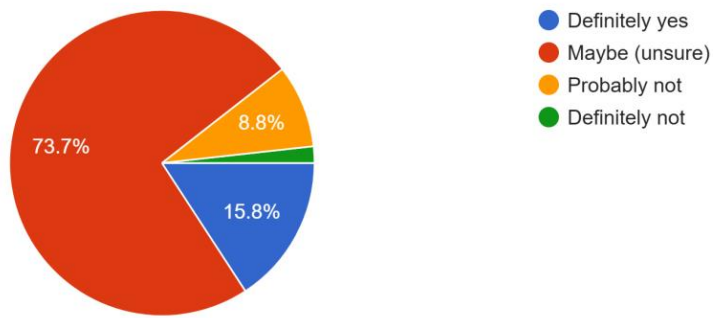

- Definitely yes — 15.8%
- Maybe (unsure) — 73.7%
- Probably not — 8.8%
- Definitely not

If you do not currently use (or plan to use) an EV, what is the main reason? (Select one.)
57 responses

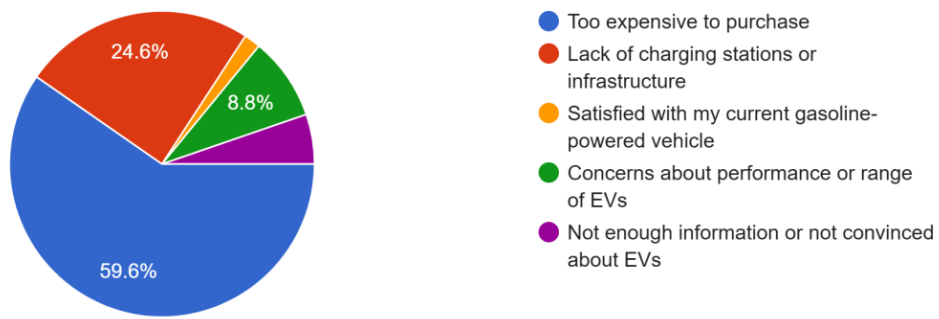

- Too expensive to purchase — 59.6%
- Lack of charging stations or infrastructure — 24.6%
- Satisfied with my current gasoline-powered vehicle
- Concerns about performance or range of EVs — 8.8%
- Not enough information or not convinced about EVs

Which of the following do you consider as potential benefits of electric vehicles? (Select all that apply.)
57 responses

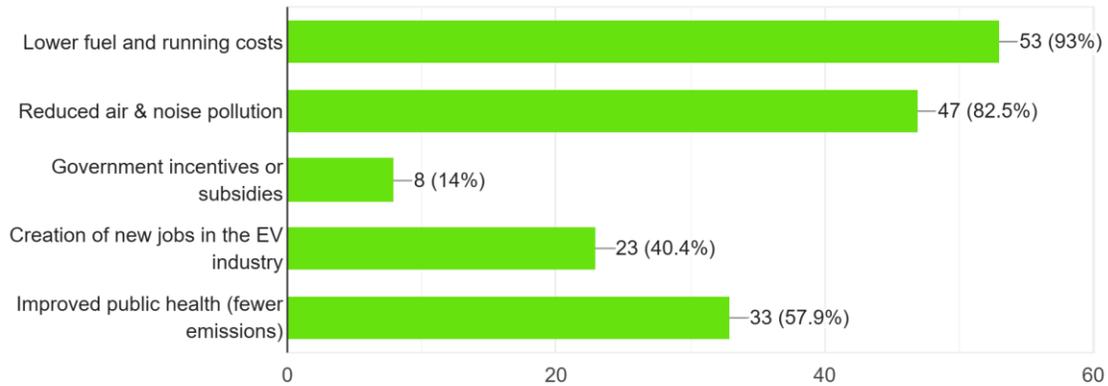

To what extent do you agree with the following statement? "Electric vehicle adoption will create new job opportunities in industries like manufacturing and maintenance."
57 responses

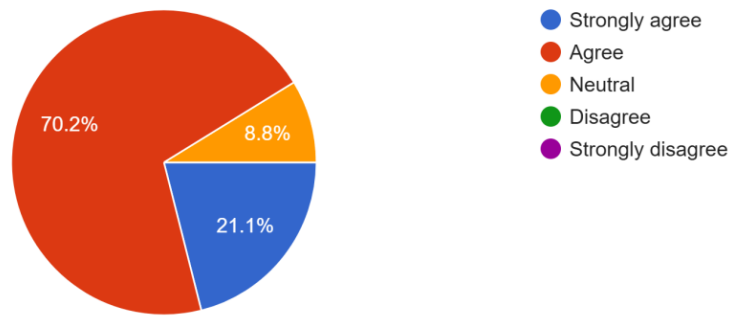

To what extent do you agree with the following statement? "Electric vehicles help reduce urban air pollution and noise compared to gasoline vehicles."
57 responses

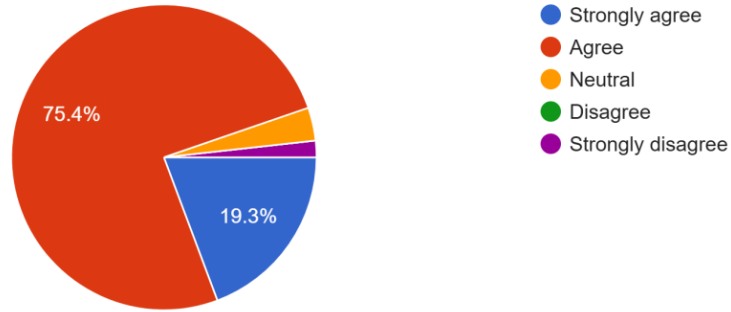

- Strongly agree
- Agree
- Neutral
- Disagree
- Strongly disagree

75.4%
19.3%

Would you be willing to pay a higher purchase price for an electric vehicle if it meant lower fuel costs and environmental benefits?
57 responses

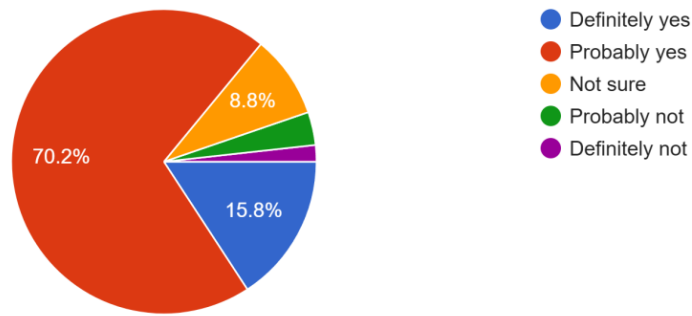

- Definitely yes
- Probably yes
- Not sure
- Probably not
- Definitely not

70.2%
15.8%
8.8%

Are you aware of any government initiatives or policies in Bangladesh that support the adoption of electric vehicles (EVs)?
57 responses

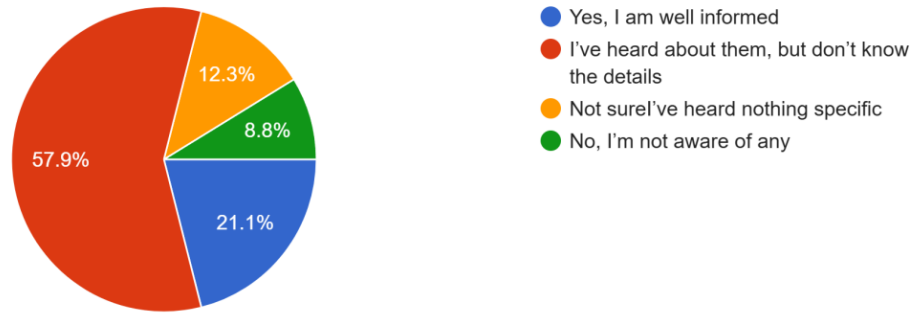

- Yes, I am well informed: 21.1%
- I've heard about them, but don't know the details: 57.9%
- Not sure/I've heard nothing specific: 12.3%
- No, I'm not aware of any: 8.8%

Do you agree or disagree: "The government should reduce taxes or provide subsidies to encourage electric vehicle adoption in Bangladesh."
57 responses

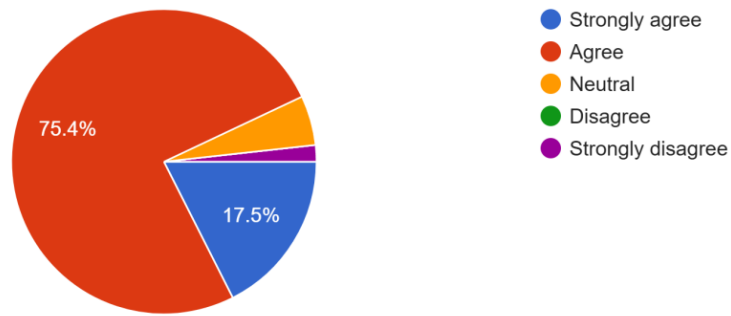

- Strongly agree: 17.5%
- Agree: 75.4%
- Neutral
- Disagree
- Strongly disagree

Do you believe electric vehicles can improve transportation accessibility for lower-income groups (e.g., e-rickshaws or shared EV rides)?
57 responses

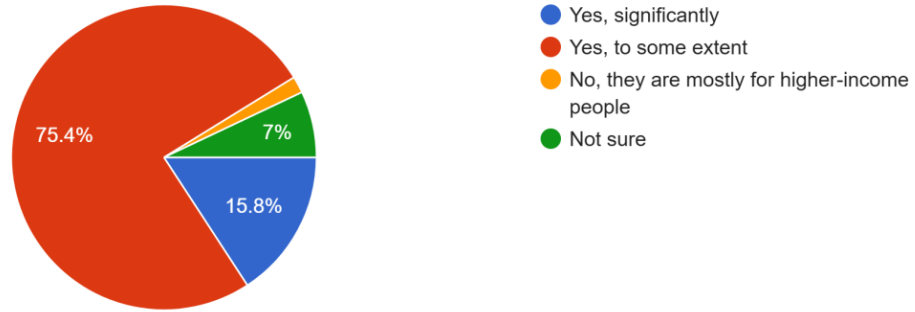

- Yes, significantly
- Yes, to some extent
- No, they are mostly for higher-income people
- Not sure

- 75.4%
- 7%
- 15.8%

In your opinion, what is the biggest obstacle to EV adoption in Bangladesh? (Select one)
57 responses

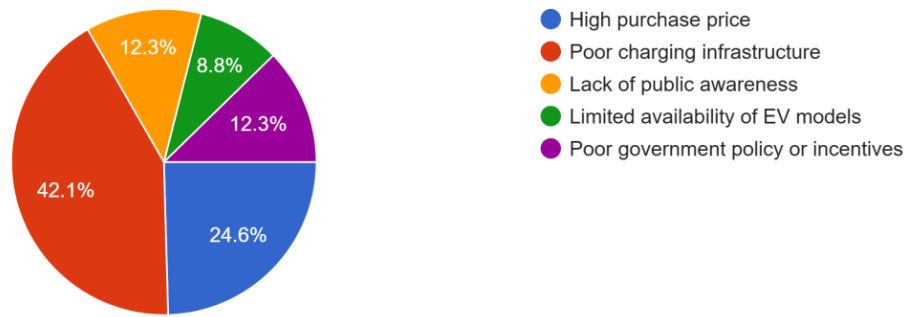

- High purchase price
- Poor charging infrastructure
- Lack of public awareness
- Limited availability of EV models
- Poor government policy or incentives

- 12.3%
- 8.8%
- 12.3%
- 42.1%
- 24.6%

Overall, what is your general attitude toward electric vehicles (EVs)?
57 responses

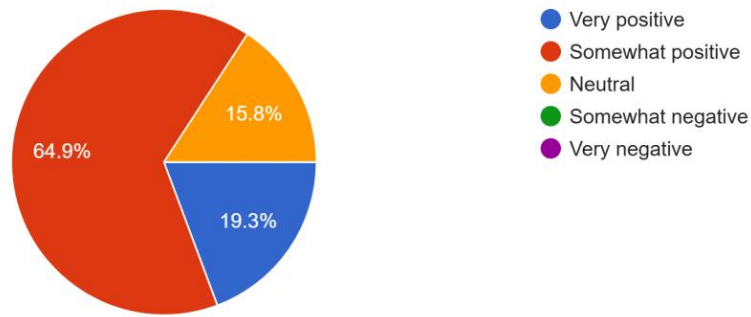

- Very positive
- Somewhat positive
- Neutral
- Somewhat negative
- Very negative

- 64.9%
- 15.8%
- 19.3%

# Analysis

According to the survey's findings, e-rickshaws which are popular and frequently used, particularly for short-distance travel—represent the primary mode of EV use in Bangladesh. This shows that EV adoption is already occurring, but in a limited capacity because most people are unfamiliar with other EV models, such as electric cars or trucks. The broader economic and environmental advantages that EVs could provide may be limited by this lack of variety.

A societal obstacle is highlighted by the information gap (28.3% having no knowledge of EV technology); without public awareness, adoption of EVs will be difficult. However, the widespread awareness of e-rickshaws indicates that EVs may expand quickly when they are accessible, affordable, and fulfil local demands.

Economically speaking, the majority of respondents (93%) believed that EVs were less expensive to operate, indicating significant long-term cost savings potential. But there are still significant obstacles, such as high buying costs (24.6%) and inadequate infrastructure (59.6%). EVs have the potential to lower transportation costs and generate employment in manufacture, maintenance, and charging services if these problems are resolved.

Regarding their effects on the environment, the high percentage of respondents (82.5%) who think that EVs lower pollution indicates that people strongly support their contribution to better air quality. If Bangladesh extends the usage of EVs beyond small passenger cars to include public buses and goods, which have a significant pollution impact, the environmental advantages will be larger.Regarding their effects on the environment, the high percentage of respondents (82.5%) who think that EVs lower pollution indicates that people strongly support their contribution to better air quality. If Bangladesh extends the usage of EVs beyond small passenger cars to include public buses and goods, which have a significant pollution impact, the environmental advantages will be larger.

Socially, EVs have the potential to make transport more inclusive, as seen by the 75.4% of respondents who believe they might increase mobility for those with low incomes. These advantages might not yet be felt by all segments of society, though, due to the present emphasis on cities and a lack of rural viewpoints.

Lastly, there is a disconnect between public communication and policy, as seen by the result that 57.9% of respondents are not aware of government EV policies. Although 75.4% of people support tax cuts or subsidies, these policies won't be successful unless the public is aware of and believes in them.

All things considered, the trends indicate that EVs are already a part of Bangladesh's transport system, but only in a limited, low-cost form. Greater awareness, improved infrastructure, and more transparent government action will be required to achieve greater economic, social, and environmental benefits.

## Economic Impacts

The adoption of electric vehicles (EVs) in Bangladesh presents significant economic opportunities, such as lower fuel imports, consumer cost savings, and the creation of jobs, there are difficulties as well, such as infrastructure development and upfront expenses.

Bangladesh's foreign exchange reserves are under a lot of strain because it now spends about $4 billion a year on gasoline imports [4]. A significant amount of this spending is attributed to the transportation industry. Making the switch to electric vehicles (EVs) could significantly reduce reliance on imported fossil fuels, especially in high-usage sectors like public transit (e-rickshaws, buses). According to a Bangladesh Power Development Board research, fuel costs might be reduced by $800 million annually if 20% of gasoline-powered vehicles were swapped out for electric vehicles [5]. In addition to improving the nation's trade balance, this change would increase energy security by reducing dependency on changes in the oil price globally.

EV adoption is strongly supported by their operational economics. According to survey results, 93% of participants are aware that EVs are less expensive to operate than traditional cars. Compared to typical autorickshaws (CNG), e-rickshaws have significantly reduced operating costs per kilometre, making them the dominant EV market in Bangladesh [6]. For commuters and drivers with modest incomes, who make up the majority of public transport users, these savings are particularly noteworthy. With 24.6% of survey participants citing cost as their top worry when it comes to EV adoption, the high initial purchase price is still a significant obstacle.

There is a lot of room for job creation in the developing EV industry across several value chains. Local manufacture (especially for e-rickshaws and battery assembly), maintenance and repair services, and the construction of charging infrastructure all present opportunities. According to a World Bank assessment, Bangladesh may add more than 50,000 new jobs in the EV industry by 2030 [7] given the right policy support. This is consistent with survey results showing that 70.2% of participants thought that the adoption of EVs would lead to job opportunities. The expansion of domestic EV industries may also boost associated industries like battery recycling and renewable energy (for charging infrastructure).

## Social Impacts

In Bangladesh, the shift to electric vehicles (EVs) is changing social dynamics by impacting community well-being, livelihoods, and public perception.

The emergence of e-rickshaws has disrupted established transportation industries and given drivers new sources of revenue. Due to cheaper operational expenses, 68% of e-rickshaw drivers reported higher incomes than traditional rickshaw pullers, according to a 2023 poll [8]. This change has, however, also resulted in conflicts with auto-rickshaw unions and sporadic demonstrations against the use of EVs in large cities. Although skill gaps continue to be a problem for workers switching from fossil fuel vehicles, the EV industry is creating jobs in maintenance workshops and battery changing stations.

As evidenced by the fact that 64.9% of poll participants had favourable opinions on electric mobility, public acceptance of EVs is increasing. Due to their reasonable charges, which are usually 20–30% less expensive than those of autorickshaws, e-rickshaws have significantly increased access to transportation for low-income people [9]. Range concern is cited by 42.1% of non-users as a barrier to adoption, indicating that there are still misconceptions regarding EV dependability. Because e-rickshaws operate more quietly and pose fewer dangers of harassment than crowded buses, female passengers say they feel safer riding in them.

Community and Safety Adoption of EVs has resulted in varying safety consequences. E-rickshaws' slower speeds (25–30 km/h) have decreased the severity of accidents in cities; since 2020, Dhaka has had an 18% decrease in fatal rickshaw accidents [10]. However, electrical fires have been caused by slums' improvised charging systems and poor battery handling.

# Environmental Impacts

Electric vehicles (EVs) can significantly reduce air pollution and greenhouse gas (GHG) emissions in Bangladesh, where traffic congestion and fossil fuel dependence worsen urban air quality.

Even though coal accounts for 10% and natural gas for 60% of Bangladesh's electricity, EVs nevertheless emit up to 30% less $CO_2$ per kilometre than gasoline or diesel cars [11]. Switching to EVs could improve public health in Dhaka by reducing dangerous pollutants by 20–25%, as $PM_{2.5}$ levels frequently surpass WHO standards [12].

Another serious problem in densely populated areas like Dhaka and Chittagong is noise pollution. Due to their near-silent operation, EVs can reduce traffic noise by 3–5 dB, potentially improving the liveability of metropolitan environments [13]. There are still issues, though, such as Bangladesh's erratic electrical system and dearth of facilities for charging. Long-term sustainability of EVs may be enhanced by the testing of solar-powered charging stations [14].

There are differing opinions among the public; many are concerned about expensive upfront expenses and battery waste. Although Bangladesh does not have enough facilities for recycling batteries, efforts are being made to securely handle lithium-ion batteries [15]. EVs have the potential to significantly reduce pollution and help Bangladesh achieve its climate goals with the right regulations.

## Policy Recommendations

The government of Bangladesh should put supportive policies into place to speed up the adoption of electric vehicles (EVs). First, EVs and their components would become less costly if high import taxes (now 25–45%) were reduced [16]. Second, it's imperative to increase the infrastructure for charging; putting in solar-powered stations along highways and cities can help with power outages [17]. Third, establishing laws governing battery recycling will stop discarded lithium-ion batteries from harming the environment [18].

Tax advantages and subsidies for local EV manufacture can increase domestic output and generate employment [19]. In order to change consumer choices, public awareness initiatives should also emphasise long-term economic savings and environmental benefits [20]. By taking these actions, Bangladesh could reduce its reliance on fossil fuels and move towards cleaner transportation.

## Conclusion

Electric vehicles (EVs) offer Bangladesh a game-changing chance to solve its increasing issues with energy security, pollution, and urban living conditions. Even with Bangladesh's present energy mix, this analysis shows that EVs can drastically lower air pollution and greenhouse gas emissions when compared to conventional automobiles. EVs are perfect for densely populated cities like Dhaka and Chittagong because of their advantages in noise reduction.

Adoption can be accelerated by specific policy measures, like as tax reductions, investments in renewable-powered charging stations, and battery recycling programs, even while obstacles like high upfront prices, a lack of charging infrastructure, and battery disposal issues still exist. EVs have the potential to significantly contribute to Bangladesh's sustainable transportation future by lowering reliance on fossil fuels and enhancing public health if properly implemented.

In addition to being a requirement for the environment, the shift to electric vehicles offers economic opportunities by promoting the expansion of local businesses and energy resilience. Bangladesh can create a cleaner, quieter, and more sustainable future by adopting EVs now.

# Appendix 1: References

[1] Ministry of Power, Energy and Mineral Resources, "EV infrastructure roadmap for Bangladesh," Government of Bangladesh, Dhaka, 2024.

 [2] S. Rahman and M. Karim, "Impact of electric vehicles on public health in urban Bangladesh," International Journal of Sustainable Transport, vol. 17, no. 4, pp. 215–229, 2024.

 [3] Bangladesh Department of Environment, "Air quality improvement through electric mobility," DOE Report, Dhaka, 2023.

[4] Bangladesh Bank, Annual Economic Review 2023, Dhaka, 2024.
[5] BPDB, Feasibility Study on EV Integration, 2023.
[6] M. R. Islam, "Cost-Benefit Analysis of E-Rickshaws in Dhaka," IEEE Access,

vol. 11, pp. 10245–10255, 2023.
[7] World Bank, Green Jobs in Bangladesh: EV Sector Potential, 2023.

[8] R. Hassan, "Income Analysis of E-Rickshaw Drivers in Dhaka," Transportation Research D, vol. 45, pp. 112-120, 2023.
[9] Bangladesh Bureau of Statistics, Urban Transport Affordability Report, 2024.
[10] Dhaka Metropolitan Police, Road Safety Annual Report, 2023.

[11] Bangladesh Power Development Board (BPDB), Annual Report 2023, Dhaka, Bangladesh, 2023.
[12] S. Rahman et al., "Air quality benefits of EVs in Dhaka," J. Environ. Manage., vol. 330, Feb. 2024, Art. no. 117456.
[13] M. Ahmed and F. Khan, "Noise reduction potential of EVs in urban Bangladesh," IEEE Trans. Sustain. Energy, vol. 14, no. 2, pp. 1023–1030, Apr. 2023.
[14] A. Hossain, "Solar-powered EV charging in Bangladesh," Renew. Energy, vol. 210, pp. 45–52, Jun. 2023.
[15] Department of Environment (DoE), E-Waste Management Guidelines, Govt. of Bangladesh, 2023.

[16] National Board of Revenue (NBR), Customs Tariff 2023, Dhaka, Bangladesh, 2023.
[17] M. Islam et al., "Solar-based EV charging solutions for Bangladesh," IEEE Access, vol. 11, pp. 45672–45683, May 2023.
[18] Department of Environment (DoE), Battery Waste Management Policy Draft, Dhaka, Bangladesh, 2024.
[19] Bangladesh Investment Development Authority (BIDA), Automotive Industry Incentives, 2023.
[20] A. Khan and R. Alam, "Consumer attitudes toward EVs in Bangladesh," Energy Policy, vol. 175, Dec. 2023, Art. no. 113478.

**ACKNOWLEDGMENT**

The author would like to thank **Prof Muhammad Abdul Awal**, Department of Electrical and Computer Engineering, North South University (NSU), for his valuable guidance and supervision throughout this research.